\documentclass[preprint,preprintnumbers,amsmath,amssymb]{revtex4}

\usepackage{graphicx}
\usepackage{dcolumn}
\usepackage{bm}

\begin{document}


\title{Shear Excitation of Confined Colloidal Suspensions} 

\author{Itai Cohen}
   \affiliation{Department of Physics and Division of
Engineering and Applied Sciences,Harvard University,9 Oxford St., Cambridge, MA 02138, USA}
\author{Thomas G. Mason}%
 \affiliation{Department of Chemistry 
and Biochemistry and Department of Physics and Astronomy,
University of California at Los Angeles, 607 Charles
E. Young Dr., Los Angeles, CA 90095, USA}
\author{David  A. Weitz}
 \affiliation{Department of Physics and Division of
Engineering and Applied Sciences,Harvard University,9 Oxford St., Cambridge, MA 02138, USA}
\begin{abstract}
We show that geometric confinement dramatically affects the 
shear-induced configurations of dense
mono-disperse colloidal suspensions; a new structure emerges, where 
layers of particles buckle to stack in a more efficient packing.
The volume fraction in the shear zone is controlled by a balance
between the viscous stresses and the osmotic pressure of a contacting
reservoir of unsheared particles. We present a model that accounts
for our observations and helps elucidate the complex interplay between
particle packing and shear stress for confined suspensions.
\end{abstract}

\maketitle
Colloidal suspensions in thermodynamic equilibrium exhibit fascinating
phase behavior, controlled by a delicate interplay between inter-particle
interactions and volume exclusion. Packing constraints are essential for determining 
the structures that result and an understanding of these
has been instrumental in elucidating the phase behavior formed by colloidal
suspensions under various conditions. These include the packing of
spherical particles in bulk \cite{Ilett1995,Yethiraj2003}, and the unavoidable 
defects formed when particles order on the surface of a spherical drop \cite{Bausch2003}.
Exposing such suspensions to large strains can drive
them out of equilibrium and significantly modify the
particle configurations; nevertheless, the shear stresses can still
effectively thermalize the particles, allowing them to explore their phase
space, and adopt reproducible structures. For example, in bulk, 
a dense suspension of mono-disperse particles subjected
to oscillatory shear will order into hexagonally-close-packed (HCP) layers 
oriented parallel to the shearing plates
\cite{Clark1979,Ackerson1990,Dux1996,Haw1998}. 
The shear-induced viscous stresses force adjacent layers to
separate allowing them to flow over one another, and the particle velocity
and the amplitude of the motion vary linearly between the shearing plates
\cite{Clark1979,Ackerson1990,Dux1996,Haw1998}.
While these structures have been well described, the complex interplay 
between particle packing and the shear induced stresses which leads to 
formation of these structures is still poorly understood. 

When a dense suspension of mono-disperse particles is geometrically confined
between two plates but not subjected to shear, the packing constraints
force the suspension to adopt equilibrium structures different from those 
observed in bulk \cite{Pieranski_pa1983_1,Van_Winkle1986,Weiss1995,Neser1997}. Subjecting 
such highly confined suspensions to shear is of considerable technological 
relevance to coatings, lubricants, and bio-rheology
\cite{Copley1963,Shaw1980,Israelachvili1991}; moreover, the limited number
of particle layers may facilitate a more quantitative analysis of the
stresses, and a direct determination of the interplay between packing and
shear-induced stresses, allowing the resultant non-equilibrium structures
to be explicitly determined. Surprisingly, such highly confined
suspensions subjected to shear have never been investigated.

In this paper, we investigate dense colloidal suspensions highly confined
between two flat plates and subjected to large oscillatory shear. We show 
that confinement forces the suspension
to adopt structures that are translationally invariant along the direction
of particle motion, and include striking gaps in the packing which, 
nevertheless, allow the particles to pack more efficiently than those 
observed in bulk.
We present a model that accounts for our observations by elucidating
the interplay between shear stress, particle packing, and geometric
confinement that leads to these ordered, but highly non-equilibrium 
structures.
\begin{figure*}[tb]{\includegraphics[clip=,height=3.1cm]{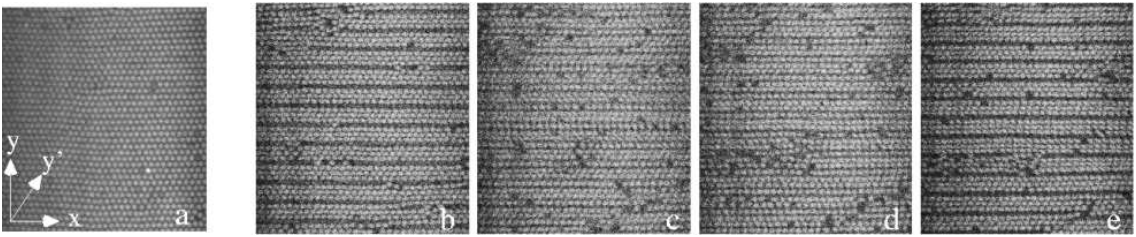}}
\caption{\label{pics}Structure of a sheared suspension 
with $\gamma = 0.38$, $f = 30$Hz, 
and $\phi = 0.61$. The plate moves in the x direction. Figure 1a shows a
confocal micrograph of a sheared suspension forming HCP layers when $D
= 80 \mu$m. Figures 1b-e show micrographs of the suspension in the
buckled state. The gap is set slightly below the height commensurate 
with confinement of four flat HCP layers. Figure 1b shows an x-y 
image slice of the suspension
near the upper plate. Figures 1c, 1d, and 1e show slices that are,
respectively, $1.3 \mu$m, $2.6 \mu$m, and $3.9 \mu$m below the slice
in 1b. The images are presented side by side so that the strip alignment 
can be compared. The y' direction is aligned with one of the characteristic 
HCP lattice vectors and forms a 60 degree angle with the x direction. 
The x, y, and y' directions are indicated in the bottom left corner
of 1a.}
\end{figure*}

The suspensions contain spherical poly-(methylmethacrylate)
particles sterically stabilized by a thin layer of poly-12-hydroxystearic
acid \cite{Weeks2000,Dinsmore2001,Antl1986}. The particles
are impregnated with rhodamine dye and suspended in a mixture of
cyclohexyl bromide, tetralin, and decalin, chosen to match both the
index of refraction and density of the particles. This procedure
allows us to view the three dimensional structure of the suspension
using fluorescence and confocal microscopy \cite{Dinsmore2001}. Optical 
tweezers measurements \cite{Crocker1999} and electrophoretic 
mobility measurements show 
that the dying process imparts a positive charge onto the particles. 
The liquid-crystal coexistence regime for these particles is 
shifted and has been shown to occur at a particle volume fraction 
$0.38 < \phi < 0.42$
\cite{Gasser2001} whereas in hard spheres, this regime occurs at
$0.494 < \phi < 0.545$. While the experiments described 
in this paper use charged particles, preliminary experiments using 
particles that are significantly less charged reproduce the observed 
phenomena and indicate that charging effects play a secondary role in 
the observed pattern formation. The solvent mixture viscosity, $\eta_0 = 0.023$poise. 
The particles have a diameter $d=1.42 \mu$m and polydispersity of $5\%$.

Our apparatus \cite{Cohen_Cell} mounts on an inverted microscope and allows 3-D 
visualization of a suspension with control over
gap separation, shear frequency, and shear amplitude. The 
suspension is sheared between a fixed plate $5$mm in diameter and a 
movable microscope cover slip, both of which are flat on the particle
length-scale. Set screws fix the gap, $D$, between $1 \mu$m and $100 \mu$m, 
and align the plates so they are parallel to
within one micron over the shear zone. A sinusoidally driven
piezoelectric actuator moves the cover slip and produces
up to $30 \mu$m displacements at frequencies, $f$, ranging between
$0$Hz and $100$Hz. The $5$mm plate is immersed in the suspension so 
particles in the shear zone contact a large reservoir of particles 
fixed at volume fraction, $\phi=0.61\pm0.02$. The cell is 
enclosed so no solvent evaporates. This lets us work with 
the suspensions for periods longer than a year.

To drive the suspension out of equilibrium, we shear with maximum strains, 
$\gamma \gtrsim 0.3$ and frequencies, $f \gtrsim 5$Hz. A phase diagram 
of the transition from 
equilibrium to nonequilibrium structures as a function of $\gamma$ and $f$ 
will be shown elsewhere\cite{Cohen_Weitz_tobe}. When the gap holds 
more than 11 layers, the morphology is 
identical to that in bulk suspensions (Fig.~\ref{pics}a). However, when the gap holds 
fewer than 11 layers, confinement plays a critical role. The HCP layer
structure becomes intermittent,
occurring only at discrete plate separations. For gaps incommensurate with 
these separations,
we observe a remarkable new ordering (Figs~\ref{pics}b-e). Fluid voids appear
within the planes and the layers break up into HCP strips aligned in
the direction of plate motion, x. Moreover, the strip widths vary with
depth, z. A typical example obtained for a gap 
slightly smaller than that confining four flat
layers, is shown by the series of x-y images in Figs~\ref{pics}b-e. Near the
stationary upper plate, the strips are three particles wide and have fluid
voids between them (Fig.~\ref{pics}b). In a plane $1.3 \mu$m lower, particles
orient in two-particle-wide strips, alternating with one-particle-wide
strips (Fig.~\ref{pics}c). Remarkably, the velocity of the two-particle-wide strips 
is larger than that of the one-particle-wide strips. This structure
repeats $1.3 \mu$m further down, but this time the velocity of the one-particle-wide
strips is larger (Fig.~\ref{pics}d). Finally, $1.3 \mu$m further down,
the layer nearest the bottom shearing plate is again oriented into
three-particle-wide strips separated by fluid voids (Fig.~\ref{pics}e). It is 
convenient to examine the strips along
the HCP lattice vector directions. Inspection of Fig.~\ref{pics}b along the y'
axis shows that the three-particle-wide strips are registered. Microscopy
measurements indicate the registration arises from interdigitation
with the one-particle-wide strips which align with the fluid voids but
are located in the plane $1.3 \mu$m below (Fig.~\ref{pics}c). This
configuration forces the one and three particle-wide strips to have
equal velocities. Similar interdigitation is observed for the
two-particle-wide strips in Figs~\ref{pics}c and~\ref{pics}d. Again, the
interdigitation forces the strips in different layers to have
equal velocities. Finally, similar behavior is observed for the one
and three particle-wide strips in Figs~\ref{pics}d and~\ref{pics}e.

\begin{figure}[b]{\centerline{\includegraphics[clip=,height=1.9cm]{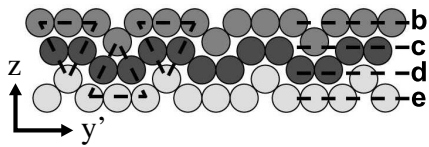}}}
\caption{\label{yz}Suspension structure in the y'-z plane.}
\end{figure}

We summarize this behavior in the y'-z schematic of the particle
positions shown in Fig.~\ref{yz}. The dashed lines indicate the z position
of the x-y planes depicted in Figs~\ref{pics}b-e. The shadings delineate
particles in contact and moving with equal velocities. This
figure shows the peculiar patterns result from a buckling of
the colloidal and lubricating fluid layers. The particle velocity 
and oscillation amplitude of the buckled particle layers 
still vary linearly between the shearing plates. Since the two sets of 
particle strips shown in figs~\ref{pics}c and~\ref{pics}d belong to different 
layers they move with different velocities. This is in sharp contrast with 
the behavior of sheared bulk suspensions where the layers always 
remain flat.

In confined unsheared suspensions, the reservoir osmotic pressure,
$\Pi_{res}$, sets the volume fraction of particles in the gap. Then,
the interplay between the particle osmotic pressure in the gap and the confined
geometry determines the suspension structure \cite{Pieranski_pa1983_1,Van_Winkle1986}. 
When the shear rate is large enough to produce non-equilibrium structures, the 
viscous stress, $\tau_{\eta}$, dominates and determines the suspension structure.
Therefore, the effective pressure arising from the shear
stress \cite{Brady1994} must balance $\Pi_{res}$ to
determine the volume fraction in the shear zone. If we start with a
commensurate configuration and increase the gap keeping the layers
flat, $\tau_{\eta}$ will decrease since there will be more fluid
between the layers. To maintain the balance between $\tau_{\eta}$ and
$\Pi_{res}$ the layers must increase their volume fraction. However, 
due to the constraints imposed by the confined geometry, sufficiently dense
packing cannot be achieved with flat HCP layers. Instead,
the system must adopt a buckled configuration. The shear stress
forces the suspension to have translational invariance along the 
direction of particle motion. Thus the suspension cannot adopt 
a configuration which corresponds to an optimized 3-D packing. Instead, 
it must optimize its packing within the y'-z plane.
Indeed, eliminating the
fluid gaps and grouping the particles into triangles (Fig.~\ref{yz}), we 
find the buckled structures resemble the
densest 2-D packing of static hard disks under incommensurate
confinement \cite{Pieranski_pi1979}. 

To estimate $\tau_{\eta}$, we consider a gap where the suspension
forms flat HCP layers. Since the shear takes place in the lubricating
fluid between the layers, the effective viscosity, $\eta_{eff}$, of
the combined structure is: $\eta_{eff}  = \eta_0 D / \sum_i l_i$,
where $l_i$ is the thickness of the $i$th lubricating fluid
layer. The viscous stress is $\tau_{\eta} \approx \gamma f \eta_{eff}$. Intriguingly, 
the HCP layer structure introduces an ambiguity in the calculation since 
different flow configurations lead to different $l_i$. Nevertheless, we 
can calculate $\tau_{\eta}$ for two limiting flow configurations. In the
straight flow configuration, particles move directly over the peaks of
the HCP sheet below. In this case, we set $l_i = 0$ when the peaks touch and
the inter-layer distance is $d$. In the zigzag flow configuration,
particles move above the valleys formed by particles in the layer
below \cite{Ackerson1990}. In this case, we set $l_i = 0$ when the interlayer
distance is $d\sqrt{3}/2$. This is the minimum separation for flow
configurations constrained to move without transverse
displacements. By measuring the interlayer spacing for the different
flows, we find that $\tau_{\eta} \approx 6.0$dyn/cm$^2$ independent of
the flow configuration.  This stress must balance $\Pi_{res}$ to
ensure no net flux of particles between the shear zone and the
reservoir. Simulations of unsheared disordered hard
spheres \cite{Woodcock1981} show $\Pi = 1.1$dyn/cm$^2$ when $\phi =
0.61$. Furthermore, in charged spheres $\Pi$ can easily reach six
times this value \cite{Phan_1996}. Thus, the stress balance is
consistent with our observations.

\begin{figure}[tb]
\includegraphics[width=0.45\textwidth]{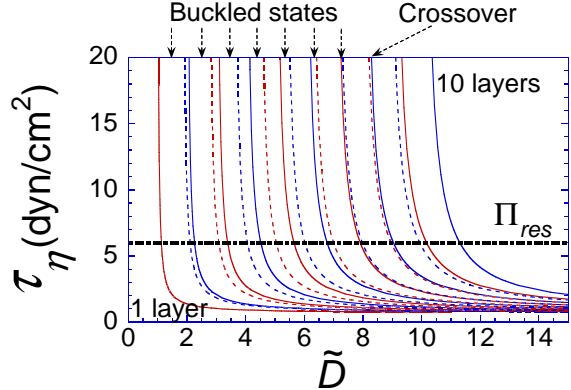}
\caption{\label{crossover}Transition mechanism predicting buckling in confined suspensions. 
The solid and dashed curves indicate
$\tau_\eta$ versus the rescaled gap, $\tilde{D}$, for systems with up
to 10 layers. The solid curves correspond to flat HCP sheets moving in
the straight flow configuration. The dashed curves correspond to the 
zigzag flow configurations and delineate the
border between the zigzag and buckled phases. With increasing gap, the
buckled region becomes smaller and eventually disappears when the
dashed and solid curves cross. The horizontal dash-dot line
corresponds to the value of $\Pi_{res}$.}
\end{figure}

This stress balance helps determine the non-equilibrium
structure of the suspension for small $D$. We define
$\tilde{D} \equiv D/d$, set $\gamma f = 30$s$^{-1}$, and, in Fig.~\ref{crossover},
plot $\tau_{\eta}$ versus $\tilde{D}$ for structures with up to $10$
layers. The solid and dashed curves correspond to $\tau_{\eta}$ values
calculated for straight and zigzag flow configurations
respectively. The horizontal dash-dot line denotes $\Pi_{res} = 6.0$
dyn/cm$^2$. At each gap, the system must assume a structure where
$\tau_{\eta} = \Pi_{res}$. Therefore, as $\tilde{D}$ is reduced, a
system initially in a straight flow configuration must increase the
amount of zigzag with which the flat HCP layers move. For systems with
fewer than eight layers, the stress curves for such configurations
always reside between a solid curve at high $\tilde{D}$ and a dashed
curve at low $\tilde{D}$ (Fig.~\ref{crossover}). However, as 
$\tilde{D}$ is reduced further, the
system must form a buckled structure with one fewer layer. Further
reduction in $\tilde{D}$ causes the amplitude of the buckles to
decrease. Eventually the layers become flat and the system is
described by the next solid stress curve with one fewer flat layers.

\begin{figure}[t]
\includegraphics[width=0.45\textwidth]{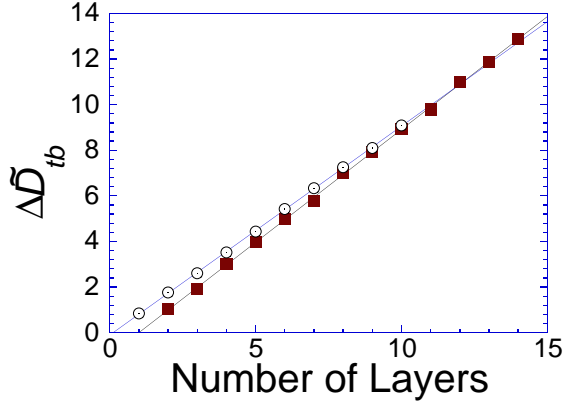}
\caption{\label{DeltaD}Separation between the top and bottom layers,
$\Delta \tilde{D}_{tb}$, for maximally buckled (open symbols) and straight 
(closed symbols) flow configurations with different numbers of layers. The measurement 
error is on the order of the symbol size. The shear rate $\gamma f = 30$s$^{-1}$. A
crossover of the curves occurs when the layer number is 12.}
\end{figure}

We can further confirm this picture by accounting for the
disappearance of the buckled state at large gaps. The $\tilde{D}$
separation between the dashed curves in Fig.~\ref{crossover} is smaller than that of
the solid curves. Consequently, the buckled state regions shrink with
increasing $\tilde{D}$ and eventually vanish when the curves cross.
The curve crossing indicates that more than one flow configuration
satisfies the pressure balance for gaps with more than 8 layers. To
investigate this crossover in the experiments, we fix $\gamma f =
30$s$^{-1}$ and plot the separation between the top and bottom layers,
$\Delta \tilde{D}_{tb}$, for the maximally buckled (open symbols) and straight
(closed symbols) flow configurations with different numbers of layers
(Fig.~\ref{DeltaD}). This measurement indicates the crossover appears when the
system reaches 12 layers. The mismatch in crossover values suggests
that in the straight flow configuration, the HCP sheets also move with some degree of
registration so that the minimum interlayer separation is less than
$d$. Indeed, even in the buckled state, where layers move with the
least amount of zigzag, ping-pong-ball-models show the minimum
interlayer separation is $0.94d$. This separation would lead to a
crossover at 13 flat layers which is in excellent accord with our
observations.

An additional test of the model is obtained by observing how the slope
of the linear fit to the buckled state data in Fig.~\ref{DeltaD} changes with
shear rate. By varying $\gamma f$, the position where the dashed
$\tau_{\eta}$ curves intersect the $\Pi_{res}$ line shifts. Therefore,
the manner in which the $\tilde{D}$ spacing between the $\tau_{\eta}$ curves
varies predicts the shear rate dependence of the slope.  For
$0.3$s$^{-1}$ $< \gamma f < 30$s$^{-1}$, the range of shear rates over which 
the measurements could be taken, the predicted slope is $0.95$ at $\gamma f = 30$s$^{-1}$ and
slowly approaches 0.87$(\approx \sqrt{3}/2)$ as the shear rate is
reduced. In the experiments, we observe a slope of $0.92 \pm 0.05$ at
$\gamma f = 30$s$^{-1}$ which gradually decreases to $0.89 \pm 0.05$
at $\gamma f = 0.3$s$^{-1}$. Again, the measurements are in excellent
agreement with the predictions.

Our experiments clearly elucidate the complex interplay between the 
reservoir osmotic pressure and the shear stress in determining the unusual 
packings of confined colloidal suspensions subjected to high shear.
Our results represent an
important instance in which the non-equilibrium configurations of a
sheared suspension can be determined. Similar effects will occur and must be considered in
other flow geometries provided the shear stress dominates in the shear
zone and the osmotic pressure dominates in the reservoir; for example 
such an interplay would be expected in a standard rheometer with a Couette 
geometry. This interplay may allow for 
tuning of the sheared suspension's rheological properties. Finally,
since the observed structures are a consequence of sphere packing in
confined geometries, such packing considerations may also impact the 
trends observed in the shear of very thin atomic and granular
films \cite{Jacob1988,Gao1997,Mueth2000}.
\begin{acknowledgments}
We thank Jacob Israelachvili, Sidney Nagel, Bruce Ackerson, David Leighton, 
John Brady, and Darren Link for helpful discussions and
Andrew Schofield and Peter Pusey for the PMMA particles. Finally, we
gratefully acknowledge the financial support of the NSF DMR-0243715, 
and NASA NAG3-2284 and NAG3-2381 grants.
\end{acknowledgments}
\noindent

\begin{thebibliography}{28}
\expandafter\ifx\csname natexlab\endcsname\relax\def\natexlab#1{#1}\fi
\expandafter\ifx\csname bibnamefont\endcsname\relax
  \def\bibnamefont#1{#1}\fi
\expandafter\ifx\csname bibfnamefont\endcsname\relax
  \def\bibfnamefont#1{#1}\fi
\expandafter\ifx\csname citenamefont\endcsname\relax
  \def\citenamefont#1{#1}\fi
\expandafter\ifx\csname url\endcsname\relax
  \def\url#1{\texttt{#1}}\fi
\expandafter\ifx\csname urlprefix\endcsname\relax\def\urlprefix{URL }\fi
\providecommand{\bibinfo}[2]{#2}
\providecommand{\eprint}[2][]{\url{#2}}

\bibitem[{\citenamefont{Ilett et~al.}(1995)\citenamefont{Ilett, Orrock, Poon,
  and Pusey}}]{Ilett1995}
\bibinfo{author}{\bibfnamefont{S.~M.} \bibnamefont{Ilett}},
  \bibinfo{author}{\bibfnamefont{A.}~\bibnamefont{Orrock}},
  \bibinfo{author}{\bibfnamefont{W.~C.~K.} \bibnamefont{Poon}},
  \bibnamefont{and} \bibinfo{author}{\bibfnamefont{P.~N.} \bibnamefont{Pusey}},
  \bibinfo{journal}{Phys. Rev. E} \textbf{\bibinfo{volume}{51}},
  \bibinfo{pages}{1344} (\bibinfo{year}{1995}).

\bibitem[{\citenamefont{Yethiraj and van Blaaderen}(2003)}]{Yethiraj2003}
\bibinfo{author}{\bibfnamefont{A.}~\bibnamefont{Yethiraj}} \bibnamefont{and}
  \bibinfo{author}{\bibfnamefont{A.}~\bibnamefont{van Blaaderen}},
  \bibinfo{journal}{Nature} \textbf{\bibinfo{volume}{421}},
  \bibinfo{pages}{513} (\bibinfo{year}{2003}).

\bibitem[{\citenamefont{Bausch et~al.}(2003)\citenamefont{Bausch, Bowick,
  Cacciuto, Dinsmore, Hsu, Nelson, Nikolaides, Travesset, and
  Weitz}}]{Bausch2003}
\bibinfo{author}{\bibfnamefont{A.~R.} \bibnamefont{Bausch}},
  \bibinfo{author}{\bibfnamefont{M.~J.} \bibnamefont{Bowick}},
  \bibinfo{author}{\bibfnamefont{A.}~\bibnamefont{Cacciuto}},
  \bibinfo{author}{\bibfnamefont{A.~D.} \bibnamefont{Dinsmore}},
  \bibinfo{author}{\bibfnamefont{M.~F.} \bibnamefont{Hsu}},
  \bibinfo{author}{\bibfnamefont{D.~R.} \bibnamefont{Nelson}},
  \bibinfo{author}{\bibfnamefont{M.~G.} \bibnamefont{Nikolaides}},
  \bibinfo{author}{\bibfnamefont{A.}~\bibnamefont{Travesset}},
  \bibnamefont{and} \bibinfo{author}{\bibfnamefont{D.~A.} \bibnamefont{Weitz}},
  \bibinfo{journal}{Science} \textbf{\bibinfo{volume}{299}},
  \bibinfo{pages}{1716} (\bibinfo{year}{2003}).

\bibitem[{\citenamefont{Clark et~al.}(1979)\citenamefont{Clark, Hurd, and
  Ackerson}}]{Clark1979}
\bibinfo{author}{\bibfnamefont{N.~A.} \bibnamefont{Clark}},
  \bibinfo{author}{\bibfnamefont{A.~J.} \bibnamefont{Hurd}}, \bibnamefont{and}
  \bibinfo{author}{\bibfnamefont{B.~J.} \bibnamefont{Ackerson}},
  \bibinfo{journal}{Nature} \textbf{\bibinfo{volume}{281}}, \bibinfo{pages}{57}
  (\bibinfo{year}{1979}).

\bibitem[{\citenamefont{Ackerson}(1990)}]{Ackerson1990}
\bibinfo{author}{\bibfnamefont{B.~J.} \bibnamefont{Ackerson}},
  \bibinfo{journal}{J. Rheol.} \textbf{\bibinfo{volume}{34}},
  \bibinfo{pages}{553} (\bibinfo{year}{1990}).

\bibitem[{\citenamefont{Dux et~al.}(1996)\citenamefont{Dux, Versmold, Reus,
  Zemb, and Lindner}}]{Dux1996}
\bibinfo{author}{\bibfnamefont{C.}~\bibnamefont{Dux}},
  \bibinfo{author}{\bibfnamefont{H.}~\bibnamefont{Versmold}},
  \bibinfo{author}{\bibfnamefont{V.}~\bibnamefont{Reus}},
  \bibinfo{author}{\bibfnamefont{T.}~\bibnamefont{Zemb}}, \bibnamefont{and}
  \bibinfo{author}{\bibfnamefont{P.}~\bibnamefont{Lindner}},
  \bibinfo{journal}{J. Chem. Phys} \textbf{\bibinfo{volume}{104}},
  \bibinfo{pages}{6369} (\bibinfo{year}{1996}).

\bibitem[{\citenamefont{Haw et~al.}(1998)\citenamefont{Haw, Poon, and
  Pusey}}]{Haw1998}
\bibinfo{author}{\bibfnamefont{M.~D.} \bibnamefont{Haw}},
  \bibinfo{author}{\bibfnamefont{W.~C.~K.} \bibnamefont{Poon}},
  \bibnamefont{and} \bibinfo{author}{\bibfnamefont{P.~N.} \bibnamefont{Pusey}},
  \bibinfo{journal}{Phys. Rev. E} \textbf{\bibinfo{volume}{57}},
  \bibinfo{pages}{6859} (\bibinfo{year}{1998}).

\bibitem[{\citenamefont{Pieranski et~al.}(1983)\citenamefont{Pieranski,
  Strzelecki, and Pansu}}]{Pieranski_pa1983_1}
\bibinfo{author}{\bibfnamefont{P.}~\bibnamefont{Pieranski}},
  \bibinfo{author}{\bibfnamefont{L.}~\bibnamefont{Strzelecki}},
  \bibnamefont{and} \bibinfo{author}{\bibfnamefont{B.}~\bibnamefont{Pansu}},
  \bibinfo{journal}{Phys. Rev. Lett.} \textbf{\bibinfo{volume}{50}},
  \bibinfo{pages}{900} (\bibinfo{year}{1983}).

\bibitem[{\citenamefont{Van~Winkle and Murray}(1986)}]{Van_Winkle1986}
\bibinfo{author}{\bibfnamefont{D.~H.} \bibnamefont{Van~Winkle}}
  \bibnamefont{and} \bibinfo{author}{\bibfnamefont{C.~A.}
  \bibnamefont{Murray}}, \bibinfo{journal}{Phys. Rev. A}
  \textbf{\bibinfo{volume}{34}}, \bibinfo{pages}{562} (\bibinfo{year}{1986}).

\bibitem[{\citenamefont{Weiss et~al.}(1995)\citenamefont{Weiss, Oxtoby, Grier,
  and Murray}}]{Weiss1995}
\bibinfo{author}{\bibfnamefont{J.~A.} \bibnamefont{Weiss}},
  \bibinfo{author}{\bibfnamefont{D.~W.} \bibnamefont{Oxtoby}},
  \bibinfo{author}{\bibfnamefont{D.~G.} \bibnamefont{Grier}}, \bibnamefont{and}
  \bibinfo{author}{\bibfnamefont{C.~A.} \bibnamefont{Murray}},
  \bibinfo{journal}{J. Chem. Phys.} \textbf{\bibinfo{volume}{103}},
  \bibinfo{pages}{1180} (\bibinfo{year}{1995}).

\bibitem[{\citenamefont{Neser et~al.}(1997)\citenamefont{Neser, Bechinger,
  Leiderer, and Palberg}}]{Neser1997}
\bibinfo{author}{\bibfnamefont{S.}~\bibnamefont{Neser}},
  \bibinfo{author}{\bibfnamefont{C.}~\bibnamefont{Bechinger}},
  \bibinfo{author}{\bibfnamefont{P.}~\bibnamefont{Leiderer}}, \bibnamefont{and}
  \bibinfo{author}{\bibfnamefont{T.}~\bibnamefont{Palberg}},
  \bibinfo{journal}{Phys. Rev. Lett.} \textbf{\bibinfo{volume}{79}},
  \bibinfo{pages}{2348} (\bibinfo{year}{1997}).

\bibitem[{Cop(1963)}]{Copley1963}
in \emph{\bibinfo{booktitle}{Proceedings of the fourth International congress
  on rheology}}, edited by \bibinfo{editor}{\bibfnamefont{A.~L.}
  \bibnamefont{Copley}} (\bibinfo{publisher}{John Wiley \& Sons, inc.},
  \bibinfo{address}{New York}, \bibinfo{year}{1963}).

\bibitem[{\citenamefont{Shaw}(1980)}]{Shaw1980}
\bibinfo{author}{\bibfnamefont{D.~J.} \bibnamefont{Shaw}},
  \emph{\bibinfo{title}{Introduction to colloid and surface chemistry}}
  (\bibinfo{publisher}{Boston: Butterworths}, \bibinfo{address}{London},
  \bibinfo{year}{1980}).

\bibitem[{\citenamefont{Israelachvili}(1991)}]{Israelachvili1991}
\bibinfo{author}{\bibfnamefont{J.~N.} \bibnamefont{Israelachvili}},
  \emph{\bibinfo{title}{Intermolecular and surface forces}}
  (\bibinfo{publisher}{San Diego Academic Press}, \bibinfo{address}{London},
  \bibinfo{year}{1991}).

\bibitem[{\citenamefont{Weeks et~al.}(2000)\citenamefont{Weeks, Crocker,
  Levitt, Schofield, and Weitz}}]{Weeks2000}
\bibinfo{author}{\bibfnamefont{E.~R.} \bibnamefont{Weeks}},
  \bibinfo{author}{\bibfnamefont{J.~C.} \bibnamefont{Crocker}},
  \bibinfo{author}{\bibfnamefont{A.~C.} \bibnamefont{Levitt}},
  \bibinfo{author}{\bibfnamefont{A.}~\bibnamefont{Schofield}},
  \bibnamefont{and} \bibinfo{author}{\bibfnamefont{D.~A.} \bibnamefont{Weitz}},
  \bibinfo{journal}{Science} \textbf{\bibinfo{volume}{287}},
  \bibinfo{pages}{627} (\bibinfo{year}{2000}).

\bibitem[{\citenamefont{Dinsmore et~al.}(2001)\citenamefont{Dinsmore, Weeks,
  Prasad, Levitt, and Weitz}}]{Dinsmore2001}
\bibinfo{author}{\bibfnamefont{A.~D.} \bibnamefont{Dinsmore}},
  \bibinfo{author}{\bibfnamefont{E.~R.} \bibnamefont{Weeks}},
  \bibinfo{author}{\bibfnamefont{V.}~\bibnamefont{Prasad}},
  \bibinfo{author}{\bibfnamefont{A.~C.} \bibnamefont{Levitt}},
  \bibnamefont{and} \bibinfo{author}{\bibfnamefont{D.~A.} \bibnamefont{Weitz}},
  \bibinfo{journal}{Appl. Opt.} \textbf{\bibinfo{volume}{40}},
  \bibinfo{pages}{4152} (\bibinfo{year}{2001}).

\bibitem[{\citenamefont{Antl et~al.}(1986)\citenamefont{Antl, Goodwin,
  Ottewill, Owens, Papworth, and Waters}}]{Antl1986}
\bibinfo{author}{\bibfnamefont{L.}~\bibnamefont{Antl}},
  \bibinfo{author}{\bibfnamefont{J.~W.} \bibnamefont{Goodwin}},
  \bibinfo{author}{\bibfnamefont{R.~H.} \bibnamefont{Ottewill}},
  \bibinfo{author}{\bibfnamefont{S.~M.} \bibnamefont{Owens}},
  \bibinfo{author}{\bibfnamefont{S.}~\bibnamefont{Papworth}}, \bibnamefont{and}
  \bibinfo{author}{\bibfnamefont{J.~A.} \bibnamefont{Waters}},
  \bibinfo{journal}{Colloids Surf} \textbf{\bibinfo{volume}{17}},
  \bibinfo{pages}{67} (\bibinfo{year}{1986}).

\bibitem[{\citenamefont{Crocker et~al.}(1999)\citenamefont{Crocker, Matteo,
  Dinsmore, and Yodh}}]{Crocker1999}
\bibinfo{author}{\bibfnamefont{J.~C.} \bibnamefont{Crocker}},
  \bibinfo{author}{\bibfnamefont{J.~A.} \bibnamefont{Matteo}},
  \bibinfo{author}{\bibfnamefont{A.~D.} \bibnamefont{Dinsmore}},
  \bibnamefont{and} \bibinfo{author}{\bibfnamefont{A.~G.} \bibnamefont{Yodh}},
  \bibinfo{journal}{PRL} \textbf{\bibinfo{volume}{82}}, \bibinfo{pages}{4352}
  (\bibinfo{year}{1999}).

\bibitem[{\citenamefont{Gasser et~al.}(2001)\citenamefont{Gasser, Weeks,
  Schofield, Pusey, and Weitz}}]{Gasser2001}
\bibinfo{author}{\bibfnamefont{U.}~\bibnamefont{Gasser}},
  \bibinfo{author}{\bibfnamefont{E.~R.} \bibnamefont{Weeks}},
  \bibinfo{author}{\bibfnamefont{A.}~\bibnamefont{Schofield}},
  \bibinfo{author}{\bibfnamefont{P.~N.} \bibnamefont{Pusey}}, \bibnamefont{and}
  \bibinfo{author}{\bibfnamefont{D.~A.} \bibnamefont{Weitz}},
  \bibinfo{journal}{Science} \textbf{\bibinfo{volume}{292}},
  \bibinfo{pages}{258} (\bibinfo{year}{2001}).

\bibitem[{Coh({\natexlab{a}})}]{Cohen_Cell}
\bibinfo{note}{I. Cohen, T. G. Mason, G. Carver, D. A. Weitz, and J. N.
  Israelachvili, to be published}.

\bibitem[{Coh({\natexlab{b}})}]{Cohen_Weitz_tobe}
\bibinfo{note}{I. Cohen and D. A. Weitz to be published}.

\bibitem[{\citenamefont{Nott and Brady}(1994)}]{Brady1994}
\bibinfo{author}{\bibfnamefont{P.~R.} \bibnamefont{Nott}} \bibnamefont{and}
  \bibinfo{author}{\bibfnamefont{J.~F.} \bibnamefont{Brady}},
  \bibinfo{journal}{J. Fluid Mech.} \textbf{\bibinfo{volume}{275}},
  \bibinfo{pages}{157} (\bibinfo{year}{1994}).

\bibitem[{\citenamefont{Pieranski and Finney}(1979)}]{Pieranski_pi1979}
\bibinfo{author}{\bibfnamefont{P.}~\bibnamefont{Pieranski}} \bibnamefont{and}
  \bibinfo{author}{\bibfnamefont{J.}~\bibnamefont{Finney}},
  \bibinfo{journal}{Acta Cryst.} \textbf{\bibinfo{volume}{A35}},
  \bibinfo{pages}{194} (\bibinfo{year}{1979}).

\bibitem[{\citenamefont{Woodcock}(1981)}]{Woodcock1981}
\bibinfo{author}{\bibfnamefont{L.~V.} \bibnamefont{Woodcock}},
  \bibinfo{journal}{Ann. N.Y. Acad. Sci.} \textbf{\bibinfo{volume}{371}},
  \bibinfo{pages}{274} (\bibinfo{year}{1981}).

\bibitem[{\citenamefont{Phan et~al.}(1996)\citenamefont{Phan, Russel, Cheng,
  Zhu, Chaikin, Dunsmuir, and Ottewill}}]{Phan_1996}
\bibinfo{author}{\bibfnamefont{S.}~\bibnamefont{Phan}},
  \bibinfo{author}{\bibfnamefont{W.~B.} \bibnamefont{Russel}},
  \bibinfo{author}{\bibfnamefont{Z.}~\bibnamefont{Cheng}},
  \bibinfo{author}{\bibfnamefont{J.}~\bibnamefont{Zhu}},
  \bibinfo{author}{\bibfnamefont{P.~M.} \bibnamefont{Chaikin}},
  \bibinfo{author}{\bibfnamefont{J.~H.} \bibnamefont{Dunsmuir}},
  \bibnamefont{and} \bibinfo{author}{\bibfnamefont{R.~H.}
  \bibnamefont{Ottewill}}, \bibinfo{journal}{Phys. Rev. E}
  \textbf{\bibinfo{volume}{54}}, \bibinfo{pages}{6633} (\bibinfo{year}{1996}).

\bibitem[{\citenamefont{Israelachvili et~al.}(1988)\citenamefont{Israelachvili,
  McGuiggan, and Homola}}]{Jacob1988}
\bibinfo{author}{\bibfnamefont{J.~N.} \bibnamefont{Israelachvili}},
  \bibinfo{author}{\bibfnamefont{P.~M.} \bibnamefont{McGuiggan}},
  \bibnamefont{and} \bibinfo{author}{\bibfnamefont{A.~M.}
  \bibnamefont{Homola}}, \bibinfo{journal}{Science}
  \textbf{\bibinfo{volume}{240}}, \bibinfo{pages}{189} (\bibinfo{year}{1988}).

\bibitem[{\citenamefont{Gao et~al.}(1997)\citenamefont{Gao, Luedtke, and
  Landman}}]{Gao1997}
\bibinfo{author}{\bibfnamefont{J.~P.} \bibnamefont{Gao}},
  \bibinfo{author}{\bibfnamefont{W.~D.} \bibnamefont{Luedtke}},
  \bibnamefont{and} \bibinfo{author}{\bibfnamefont{U.}~\bibnamefont{Landman}},
  \bibinfo{journal}{Phys. Rev. Lett.} \textbf{\bibinfo{volume}{79}},
  \bibinfo{pages}{705} (\bibinfo{year}{1997}).

\bibitem[{\citenamefont{Mueth et~al.}(2000)\citenamefont{Mueth, Debregeas,
  Karczmar, Eng, Nagel, and Jaeger}}]{Mueth2000}
\bibinfo{author}{\bibfnamefont{D.~M.} \bibnamefont{Mueth}},
  \bibinfo{author}{\bibfnamefont{G.~F.} \bibnamefont{Debregeas}},
  \bibinfo{author}{\bibfnamefont{G.~S.} \bibnamefont{Karczmar}},
  \bibinfo{author}{\bibfnamefont{P.~J.} \bibnamefont{Eng}},
  \bibinfo{author}{\bibfnamefont{S.~R.} \bibnamefont{Nagel}}, \bibnamefont{and}
  \bibinfo{author}{\bibfnamefont{H.~M.} \bibnamefont{Jaeger}},
  \bibinfo{journal}{Nature} \textbf{\bibinfo{volume}{406}},
  \bibinfo{pages}{385} (\bibinfo{year}{2000}).

\end{thebibliography}

\end{document}